# Estimating of optimal dose of PACL for turbidity removing from water

Anahita ghafoorisadatieh [1], Ebrhim Almatin[2], Mansooreh soleymani nezhad Bam [3], Amir Gholipour[4*]

1- Department of civil Engineering, Institute of higher education khazar Mahmudabad, Iran
2- Assistant Professor, Department of Civil Engineering, Khavaran Institute, Mashhad, Iran
3- Department of civil Engineering, Khavaran Institute of Higher Education, Mashhad, Iran.
4- PhD student at Environmental Engineering Department, University of Lisbon, Lisbon, Portugal

* Corresponding author Email: isa125145@isa.ulisboa.pt , tell: +98- 935512855

## Abstract
Removing suspended solids and colloids are one of the most important water treatment processes. In this research, experimental tests have been done to assess the effects of pH and different doses of PACL on turbidity removal efficiency. The optimum dose of poly aluminum chloride (PACL) for removing turbidity was also determined. Turbidity was created artificially by adding the Kaolin with six different initial turbidities between 20 to 300 NTU rested for half an hour. The results have been showed that the best removal efficiency was at the pH of natural water. The optimal doses consumption of PACL was about 5 PPM.
**Keywords:** Turbidity, Optimum dose, Poly Aluminum Chloride (PACL), pH.

## 1. Introduction
Turbidity is one of the important physical parameters of water that representing the amount of suspended solids, (Kawamura, 2000; Daneshvar, 2009). The turbidity value is proportional to the amount of colloid and suspended solids in water that cannot be deposited in usual ways. These particles are not visible but a whole set of them can scatter the light beams, (Daneshvar, 2009).
Suspended solids can settle and deposit in the tanks and cause problems. Zahabi et al. at 2018 carried out a numerical study on three rectangular reservoirs with different dimensional ratios to investigate the sedimentation in reservoirs (Zahabi et al. 2018). Suspended solids lead to disturbance in the disinfection of water with disinfectant, and act as a barrier between disinfectant material and microbial agents. Hence, water turbidity prevents the complete disinfection operation, or in other words, it increases the amount of disinfectant used. Also in drinking water, high turbidity causes intestinal diseases in human body, that it more dangerous for people who have vulnerable immune system. Bacteria and viruses attach to the surface of suspended solids and consequently lead to the pollution of water, therefore, according to the standard EPA, the allowed quantity of turbidity in drinking water is less than 1 NTU, (USEPA, 1999).

Coagulation–flocculation is a relatively simple physical -chemical technique that is commonly used for water and wastewater treatment, (Duan and Grogory, 2003; Ghafari et al., 2009). Coagulation is an event in which the charge particles in colloidal suspension (most of them involve negative charge) are neutralized by mutual collision with opposite ions and then they will be sediment as a mass ultimately (Daneshvar, 2009). It is a process for transforming colloidal particles into larger aggregates (flocs) and for absorbing dissolved organic matter on to the surface of particulate aggregates. These impurities can be removed in subsequent liquid/solid operation processes, (Li et al., 2010).

The most important coagulations of aluminum are hydrated aluminum sulfate (Alum), poly aluminum chloride (PACL), poly aluminum silicate sulfate (PASS) and poly aluminum silicate chloride (PASIC). Ferrous sulfate ($FeSO_4 \cdot 7H_2O$), ferric sulfate $Fe_2(SO_4)_3$, ferric chloride ($FeCl_3 \cdot 6H_2O$) and poly ferric sulfate (PFS) are some of ferric compounds (Olia et al. 2019), usually used for removing organic materials and alga, (Daneshvar, 2009). In surface water treatment, process of coagulation can be done by ferric chloride and aluminum sulfate (Alum). Using these materials is associated with a series of problems. Using ferric chloride in removing the turbidity is along with creating color in the water that causes brown yellowish spots on the objects. If its quantity in the treated water is more than 1 mg/L, it can lead to opacity and not pleasing taste in water, because of the conversion of $Fe^{2+}$ to $Fe^{3+}$ in the air vicinity (APHA, AWWA, WPCF, 1995; Aghapoor et al., 2009). The disadvantages of using of aluminum sulfate can be referred to as creating Calcium hardness and also creating $CO_2$ that is a factor for corrosion, (loee, 1998).

The results of comparing the operation of PACL with ferric chloride and alum have showed that flocs sedimentation rate has been higher by using poly aluminum chloride (Bani Hashemi et al., 2008).

Optimization of coagulation and flocculation in Pulau Burung landfill site leachate treatment with poly aluminum chloride and alum showed that the efficiency of the removal of all three; turbidity, color and TSS was higher by using of poly aluminum chloride than alum, with considering that, the dosage of alum was nearly five times as poly aluminum chloride, (Ghafari et al., 2009).

In recent years, PACL has been widely used in water and wastewater treatment to remove contaminants (Zhao et al., 2009; Xiaoying et al., 2009). In water treatment, in addition to turbidity removal, poly aluminum chloride has been used for the removal of other pollutants in coagulation process, such as Bisphenol A, that was made by combining acetone and phenol, (Xiaoying et al., 2009). Research has shown that PACL is efficient for the removal of heavy metals such as Zn, Cr , Pb (El Samrani et al., 2008) and in simultaneous removal of turbidity and humic acid from high turbidity stormwater, (Annadurai et al., 2004).

Some Al-based compounds or polymers such as alum, aluminum chloride, PACL and so on, are widely used as a coagulant in drinking water treatment to enhance the removal of particulate, colloidal, and dissolved substances via coagulation process, (Maria et al., 2004; Emma et al., 2006; Yang et al., 2010). PACL has been claimed by many investigators to be superior to the traditional Al-based coagulants (e.g., $AlCl_3$ and alum) in particulate and/or organic matter removal, (Duan and Gregory, 2003; Guan et al., 2006; Yan et al., 2007) such as positively charged monomers and polymers, formation of rapid and denser floc, reduced sludge (Viraraghavan and Wimmer, 1988; Gao et al., 2007) and effectiveness over a wide pH range (Rebhun and Lurie, 1993; Delgado et al., 2003; Matsui et al., 2003).

The performance of inorganic metal salt is creating instability of particles that is created due to electrical double layer compression around the fine particle and colloid. Where as polymers create this instability by surface attraction of colloid particle (Wu et al., 2009) and creating particle- polymer- particle bridge connection. So according to the cases mentioned above, polymeric metal salt like PACL, can be expected to have both instability performances simultaneously, which leads to the development, improvement and accelerating the operation of making unstable particles, so faster growing and also better removing of particles can be achieved, (Bani Hashemi et al., 2008). Currently, PACL is used as a coagulant in water treatment in several plants in Iran.

In this paper, determination of the optimal consumption of PACL is considered by using experimental tests.

## 2. Materials and Methods

In this research turbidity was created artificially by adding the kaolin, that it solubility in water is in form of suspension ($Al_2O_3 \cdot H_2O \cdot 2SiO_2$), to the samples of entrance raw water to the no. 3 water refinery of Mashhad that was supplied from the Dousti dam lake. The characterizations of raw water are mentioned in table 1. The effects of different coagulant (PACL) doses on made samples were considered in optimal pH condition by using the jar-test apparatus (JLT leaching test jar-test- VELP scientific). Their final turbidity was measured by using the turbidimeter, (2100P turbidimeter).

Optimal pH is determined for using the PACL. For this purpose, 6 samples were made in 1000 ml volume with 50 NTU turbidity and their pH was set in 2, 4, 6, 7, 8 and 9 respectively. The samples were placed in jar-test apparatus and 15 ppm coagulant was added to each beaker. The jar-test apparatus was set on rapid mixing speed of 200 rpm for 60 s, and then on slow mixing speed of 45 rpm for 20 min. The samples have to rest for 30 minutes. At the end of these experiments 10 ml sample was sampled from 3 cm below the liquid surface to determine the percent of turbidity removal.

**Table 1. Characterization of entrance raw water to the no. 3 water refinery of Mashhad**

| parameter | unit | Maximum allowable | The amount in raw water |
|---|---|---|---|
| Turbidity | NTU | <1 | About 1 NTU |
| pH | - | 7.5- 8 | 8 |
| temperature | c° | - | 20 |
| EC* | μg/cm² | - | 900 |
| TDS** | mg/lit | 500 | 450-480 |
| TSS*** | - | - | 2-3 |
| TH**** | $Caco_3$(mg/l) | 500 | 300- 320 |
| Calcium hardness | $Caco_3$(mg/l) | 300 | About 170 |
| Magnesium hardness | $Caco_3$(mg/l) | 200 | About 150 |
| Total alkalinity | $Caco_3$(mg/l) | - | 148-155 |

*Electrical conductivity, **Total dissolved solids, ***Total hardness, ****Total suspended solids

After determining the optimal pH, the samples with turbidity of 20, 50, 100, 150, 200 and 300 NTU, were made by adding the sufficient quantity of kaolin in optimal pH and then were tested by jar-test apparatus according to the mentioned levels in Table 2, (The tests were repeated three times).

At first the quantity of PACL coagulant was optimized and then the best rapid mixing speed was determined by optimal consumption PACL. For optimization of the quantity of PACL coagulant, the values of 3, 5, 7, 9, 11, 13, 15, 17, 19, 21, 23 and 25 ppm of PACL were injected to each of the samples which were made in turbidity that was mentioned in Table 2. Then they were tested in rapid mixing speed, 150 rpm, for 60 s, and after that in slow mixing speed, 45 rpm, for 20 min. After 30 min rest time, the remained turbidity in samples was measured and the quantity of optimal consumption PACL was determined.

Table 2. Levels of experiments

| Turbidity NTU | PACL Mg/lit | Rapid mixing time(s) | Rapid mixing speed(rpm) | Slow mixing time(min) | Slow mixing speed(rpm) |
|---|---|---|---|---|---|
| 20 | 3 | 60 | 150 | 20 | 45 |
| 50 | 5 | 120 | 200 | | |
| 100 | 7 | 180 | 250 | | |
| 150 | 9 | | | | |
| 200 | 11 | | | | |
| 300 | 13 | | | | |
| | 15 | | | | |
| | 17 | | | | |
| | 19 | | | | |
| | 21 | | | | |
| | 23 | | | | |
| | 25 | | | | |

## 3. Results & Discussion

The purpose of these experiments was to get the optimal condition in removing the turbidity from water in order to achieve the turbidity less than 1 NTU according to the EPA standard.

It should be noted that in water treatment process a part of turbidity was removed by filtration after the coagulation and flocculation phases. But for increasing the filtration efficiency, it was necessary to remove much turbidity in coagulation and flocculation process. So it was tried to control the turbidity of input water to the filters to some extent in order not to cause clogging filters, yet the turbidity of output water remain in the standard range. The results of this study were based on the final turbidity after coagulation and flocculation steps and before the filtration step.

According to the Figure 1, diagram of turbidity removal percentage against pH variations, it can be seen that increasing of pH from acidic to alkaline condition, causes the enhancement of the turbidity removal percentage; but the rate of increasing was slower. So the rate of increasing between pH=2 to pH=4 was in maximum range, but with pH above 4 this increasing was slow. Therefore it can be concluded that the highest turbidity removal percentage was in the range (7-8.5) that it is natural water pH.

Figure 2, represents the results of determining the optimal consumption doses of PACL as a coagulant. According to figure 2, it can be seen that in pH: (7- 8.5), by increasing the PACL from 3 ppm to 25 ppm, the amount of removal turbidity was changing, under these conditions: Time/Speed of rapid mixing: 60s/150 rpm, Time/Speed of slow mixing: 20 min/45 rpm, Rest time: 30 min.

In turbidity equivalent to 20 NTU, by increasing PACL from 3 ppm to 9 ppm, the turbidity removal percent was enhanced, but from 9 ppm doses to 15 ppm, increase of removal percentage was too little and in higher doses than 15ppm of PACL, the results showed a decrease in turbidity removal percentage. So, by increasing the amount of PACL in the sample, the possibility of contact of colloids and PACL as a coagulant is increased, then floc formation efficiency and so coagulation efficiency is enhanced to achieve its maximum, the turbidity removal percentage will be fixed by increasing in PACL until the removal efficiency begins to decrease with 1 ppm increase in PACL; because injected additional PACL, due to the reduction of the suspended solids and colloids in sample, wasn't able to make significant sedimentation floc. In other words, no floc was formed and extra PACL remained in water and so aqueous solution was reached to saturation. Therefore additional PACL caused turbidity in samples and also a decrease in the final turbidity removal percentage. This topic is applied to all samples with different initial turbidity with considering:

**i.** By increasing the initial turbidity, the efficiency of turbidity removal was enhanced, because the increase of suspended solids and colloids causes a better coagulation due to the possibility of creating large and heavy floc, hence in low doses of PACL, the efficiency of turbidity removal was getting better. In other words, in the same doses of coagulant in samples with different turbidity, the turbidity removal efficiency was enhanced with an increase in initial turbidity.

**ii.** In the samples with high initial turbidity, the decrease of turbidity removal efficiency, due to an increase in amount of PACL as a coagulant, happened in higher doses. By considering figure 2, two points can be understood: 1. Increase in turbidity from 20 NTU to 300 NTU causes the enhancement of the coagulant efficiency. 2. The rate of increasing turbidity removal percentage is low in doses more than 5 ppm. With regard to economic considerations, the best quantity of PACL as a coagulant can be approximated in 5 ppm.

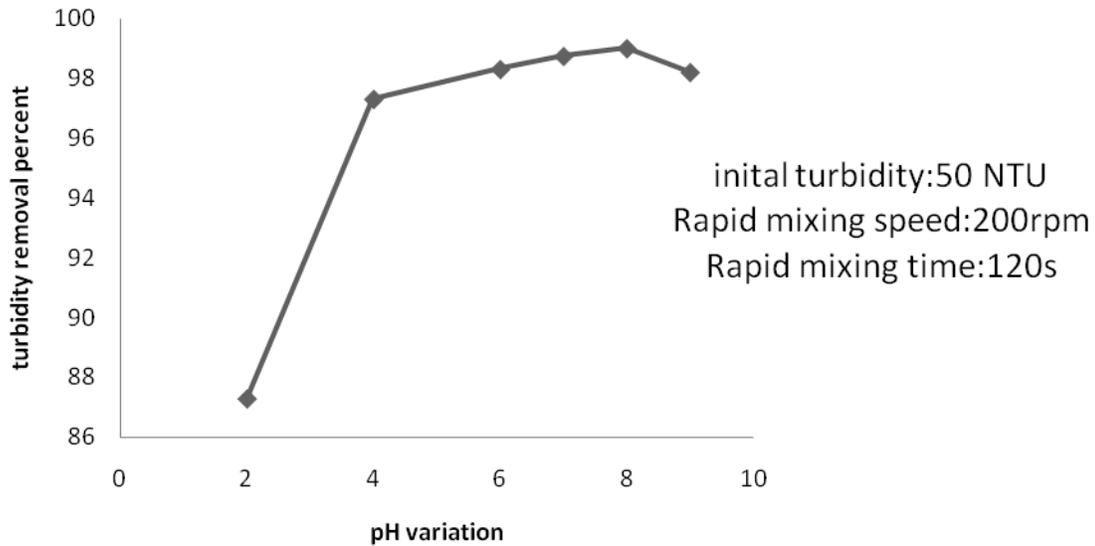

**Fig. 1: The effect of pH variation on turbidity removal percent**

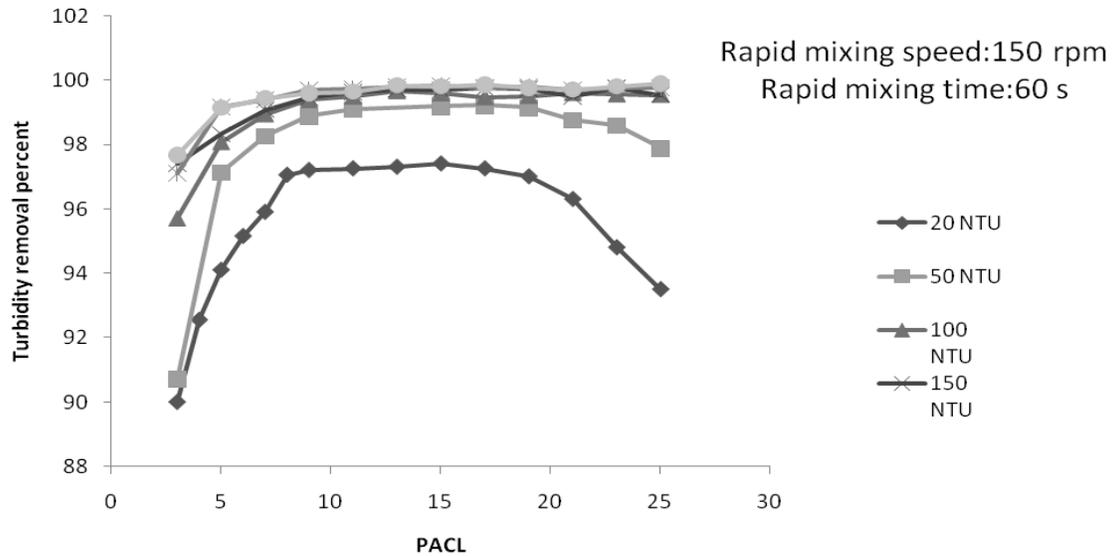

**Fig. 2:** The effect of different doses of PACL on turbidity removal percent

## 4. Conclusion
In this paper, the effects of pH and different doses of PACL on turbidity removal efficiency were studied experimentally. The efficiency of PACL operation in coagulation process, for removing the turbidity, was directly related to initial turbidity quantity and pH. The turbidity removal efficiency in higher turbidity was increased. PACL as a coagulant operated well in vast domain of pH. The coagulant efficiency became better in higher pH. Optimal pH for removing the colloids and suspended particles was about natural pH of water (7- 8.5). In turbidity less than 300 NTU, it was possible to reach turbidity removal efficiency more than 95% approximately, by using of 5 ppm of poly aluminum Chloride coagulant.

## 5. Acknowledgement
The support of the research council of Shahid Chamran University of Ahvaz, support no. 2 water refinery laboratory of Ahvaz and the no. 3 water refinery laboratory of Mashhad during the preparation of this work is kindly acknowledged.

## 6. Reference
Aghapoor, A. A., Mohamadi Boinee, A. and Hassan Zadeh, Y., (2009). Performance study of PACL as a coagulants in Chay river of Urmia. Twelfth conference of environmental health. Iran. Shahid Beheshti university of medical science. Faculty of Health., 1186- 1193 (8 pages).

Annadurai, G., Sung, S. S. and Lee, D. J., (2004). Simultaneous removal of turbidity and humic acid from high turbidity stormwater. Adv. Environ. Res., 8, 713-725.

APHA, AWWA, WPCF., (1995) . Standard Methods for the Examination of Water and Wastewater , 20th. Ed. APHA, N. W, Washington D. C.

Bani Hashemi, a., Alavi Moghadam, M. R., Macnoon, R. and Nic Azar, M., (2008). Laboratory study of inorganic polymer in water turbidity removal. Water and wastewater journal., 66, 82- 86 (5 pages).

Daneshvar, N. A., (2009). Chemical water and wastewater treatment. Amidy publisher Inc, Tabriz, Iran, 121-156, 215-268.

Delgado, S., Diaz, F., Garcia, D. and Otero, N., (2003). Behaviour of inorganic coagulants in secondary effluents from a conventional wastewater treatment plant. Research Article., 42-45.

Duan, J. and Grogory, J., (2003). Coagulation by hydrolyzing metal salts. Adv. Colloid Interface Sci., 100- 102, 475-502.

El Samrani, A. G., Lartiges, B. S. and Villieras, F., (2008). Chemical coagulation of combined sewer overflow: Heavy metal removal and treatment optimization. Water. Res., 42, 951-960.

Emma, S. L., Simon, P. A. and Bruce, J., (2006). Seasonal variations in natural organic matter and its impact on coagulation in water treatment. Sci. Total Environ., 363, 183-194.

Gao, B. Y., Yue, Q. Y. and Wang, Y., (2007). Coagulation performance of polyaluminum silicate chloride (PASiC) for water and wastewater treatment. Separation and Purification. Technol., 56, 225-230.

Ghafari, S., Aziz, H. A. and Isa, M. H., (2009). Application of response surface methodology (RSM) to optimize coagulatin- flocculation treatment of leachate using poly- aluminum chloride (pac) and alum. Hazardous Materials., 163, 650-656.

Guan, X. H., Chen, G. H. and Shang, C., (2006). Combining kinetic investigation with surface spectroscopic examination to study the role of aromatic carboxyl groups in NOM adsorption by aluminum hydroxide. Colloid and Interface. Sci., 301, 419-427.

Kawamura, S., (2000). Integrated design and operation of water treatment facilities, John Wiley & Sons Inc.

Li, F., Jiang, J. Q., Wu, S. and Zhang, B., (2010). Preparation and performance of a high purity poly-aluminum chloride. Chemical. Eng., 156, 64-69.

Loee, A., (1998). Improve the quality of drinking water using powdered activated carbon. MSc. Dissertation, Tehran university of Medical science. Faculty of Health. Iran.

Maria, T., Sylwia, M. and Antoni, M. W., (2004). Removal of organic matter by coagulation enhanced with adsorption on PAC. Desalination., 161, 79-87.

Matsui, Y., Matsushita, T., Sakuma, S., Gojo, T., Mamiya, T., Suzuki, H. and Inove, T., (2003). Virus inactivation in aluminum and polyaluminum coagulation. Environ. Sci. Technol., 37(22), 5175-5180.

Olia, H., Torabi, M., Bahiraei, M., Ahmadi, M. H., Goodarzi, M., & Safaei, M. R. (2019). Application of nanofluids in thermal performance enhancement of parabolic trough solar collector: state-of-the-art. Applied Sciences, 9(3), 463.

Rebhun, M. and Lurie, M., (1993). Control of organic matter by coagulation and floc separation. Water Sci. Tech., 27(11), 1-20.

USEPA, (1999). Guidance manual turbidity provisions, Office of water and Drinking Ground water , Washington, D. C.

Viraraghavan, T. and Wimmer, C. H., (1988). Polyaluminum chloride as an alternative to alum coagulation: a case study. Proc. Canadian Soc. Siv. Eng. Annu. Conf., 480- 498.

Wu, X., Ge, X., Wang, D. and Tang, H., (2009). Distinct mechanisms of particle aggregation induced by alum and PACL: floc structure and DLVO evaluation. Colloid and Surface A; Physicochem, Eng., 347, 56-63.

Xiaoying, M., Guangming, Z., Chang, Z., Zisong, W., Jian, Y., Jianbing, L., Guohe, H. and Hongliang, L., (2009). Characteristics of BPA removal from water by PACL-$Al_{13}$ in coagulation process. Colloid and Surface. Sci., 337, 408-413.

Yan, M., Wang, D., Qu, J., He, W. and Chow, C. W. K., (2007). Relative importance of hydrolyzed Al (III) species ($Al_a$, $Al_b$ and $Al_c$) during coagulation with polyaluminum chloride : a case study with typical micro- polluted source water. Colloid and Interface. Sci., 316, 482-489.

Yang, Z. L., Gao, B. Y., Yue, Q. Y. and Wang, Y., (2010). Effect of pH on the coagulation of Al-based coagulants and residual aluminum speciation during the treatment of humid acid-kaolin syntheric water. Hazardous Materials., 178, 596-603.

Zahabi, H., Torabi, M., Alamatian, E., Bahiraei, M., & Goodarzi, M. (2018). Effects of Geometry and Hydraulic Characteristics of Shallow Reservoirs on Sediment Entrapment. Water, 10(12), 1725.

Zhao, H. Z., Liu, C., Xu, Y. and Ni, J. R., (2009). High- concentration polyaluminum chloride: preparation and effects of the Al concentration on the distribution and transformation Al species. Chemical. Eng., 155, 528-533.